# Investigation of Turbulent Transport in Non-Stabilized Plasma at the Edge of the GOLEM Tokamak

I. Nanobashvili[1], V. Svoboda[2], S. Nanobashvili[1],
G. Van Oost[3], I. Naskidashvili[1], I. Tsevelidze[1]

[1]Andronikashvili Institute of Physics, Tbilisi, Georgia
[2]Czech Technical University in Prague, Prague, Czech Republic
[3]Ghent University, Ghent, Belgium

Plasma turbulent transport was investigated at the edge of the GOLEM tokamak. For this purpose, ion saturation current fluctuations were measured by means of Langmuir probes. Plasma stabilization was not activated during the discharges under study. Non-Gaussian behavior is observed, which indicates the presence of coherent structures. Study of the radial dependence of statistical and temporal characteristics of fluctuations allows to conclude that non-elongated blob-like structures with comparable radial and poloidal size should govern the plasma transport at the edge of the GOLEM tokamak.

## Introduction

Understanding the physical nature of plasma turbulent transport is very important for controlled thermonuclear fusion, because turbulent transport reduces confinement. At the same time, it can cause strong erosion and high heat load on plasma facing components together with unwanted retention of tritium in next generation devices like ITER [1]. Turbulent transport has bursty and intermittent characteristics due to occurrence of coherent turbulent structures [2]. These structures are formed on a diffusive background and plasma density and temperature inside them are higher compared to those in the background plasma. They propagate mostly radially at a speed which is fraction of the ion sound speed. Langmuir probes are one of the most common diagnostics used for investigation of plasma turbulent transport [2-6]. When the coherent turbulent structures propagate radially outwards, they pass the Langmuir probe pin and a burst in the temporal evolution of plasma density is detected. Generally, bursty plasma transport is a quite universal phenomenon and was detected in various fusion devices like tokamaks [7-14], stellarators [15, 16], linear devices [2, 12, 17-19], reversed field pinches [20, 21] and simple magnetized toroidal devices [22, 23]. Furthermore, plasma turbulent transport exhibits a multiscale structure and is characterized by superdiffusion [24, 25]. For this reason plasma turbulent transport at the edge of fusion devices is often called anomalous transport.

The statistical and temporal characteristics of plasma turbulent fluctuations together with their radial dependence were investigated for a better understanding of plasma turbulent transport processes.

## Experimental set-up

GOLEM is small size tokamak. It was known as CASTOR tokamak and operated at the Institute of Plasma Physics, Academy of Sciences of the Czech Republic, Prague for 30 years [26]. After that it was transferred to the Faculty of Nuclear Physics and Physical Engineering, Czech Technical University in Prague and renamed GOLEM [27].

The major radius of the GOLEM tokamak is R = 0.4 m and the minor radius r = 0.1 m. The stainless steel discharge chamber is equipped with a poloidal limiter which is made of molybdenum and has the radius a = 0.085 m. The maximum toroidal magnetic field is 0.5 T, the central electron temperature is lower than 100 eV, the maximum line average density ~$10^{19}$ $cm^{-3}$ and the discharge duration is up to 25ms.

In order to investigate plasma turbulent transport in GOLEM, dedicated experiments were performed. Namely, the ion saturation current ($I_{sat}$) was measured with sampling frequency 1 MHz by means of double rake probe inserted into the discharge chamber from the bottom diagnostics port. The rake probe consists of 12 Langmuir probe pins in two radial columns with 6 pins in each. The two columns are poloidally separated by 2.5 mm and the radial distance between the probe pins in each column is also 2.5 mm. Eight discharges were performed in order to make a radial scan of edge plasma – the probe radial position changed from shot to shot. In this paper we present the results calculated from the data obtained on the left pin closest to the nose of the radial rake probe, because this data is most unperturbed by the probe shaft.

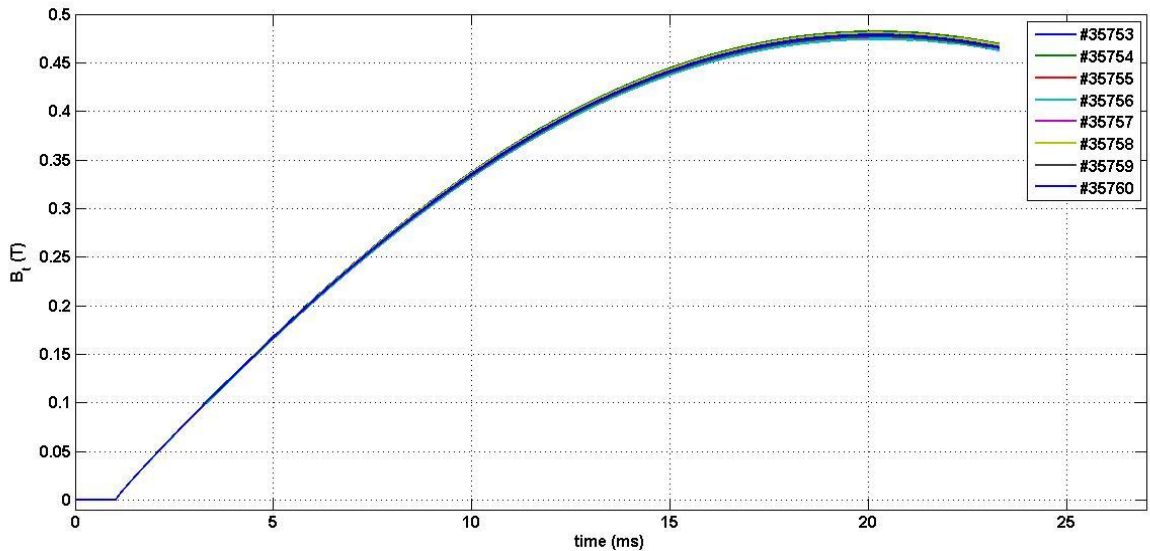

*Fig. 1 Temporal evolution of toroidal magnetic field for 8 discharges under study*

Each discharge lasted for about 25 ms and $I_{sat}$ was measured during the whole discharge. Since plasma current and toroidal magnetic field in the GOLEM tokamak increase during the discharge (see the Fig.1 and Fig.2), we analyzed the data from a 1 ms long time window after 15 ms. At this stage of discharge toroidal magnetic field and plasma current nearly reach their maximum values (around 0.45 T and 6 kA respectively) and do not change significantly (see the Fig.1 and Fig.2). Therefore, confinement should be high and machine functioning in most tokamak-like regime, namely flat-top phase of the tokamak discharge.

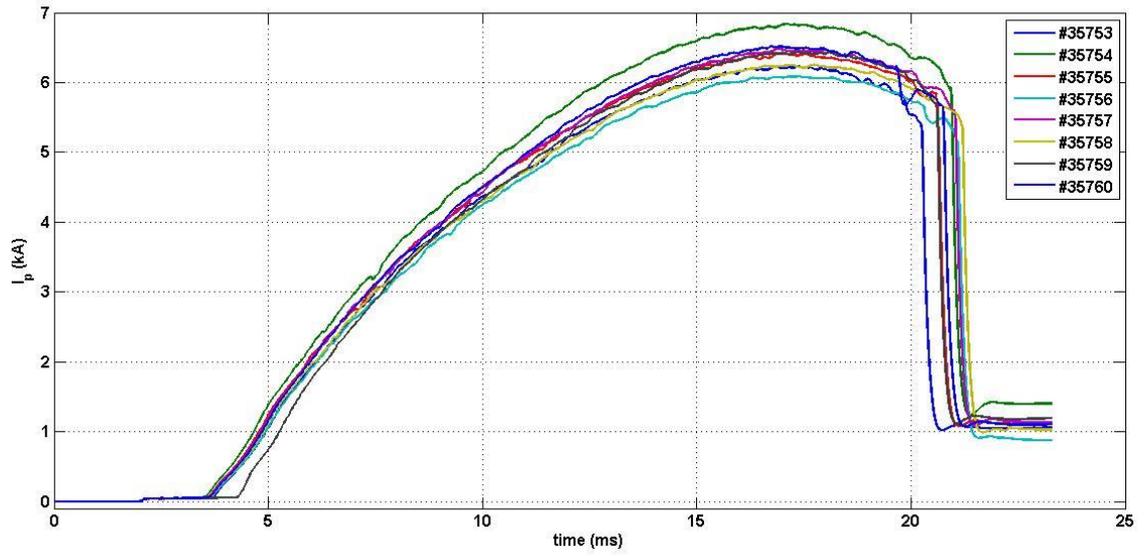

*Fig. 2* *Temporal evolution of plasma current for 8 discharges under study*

Generally, GOLEM discharges are well reproducible as shown by the temporal evolution of loop voltage (see Fig. 3).

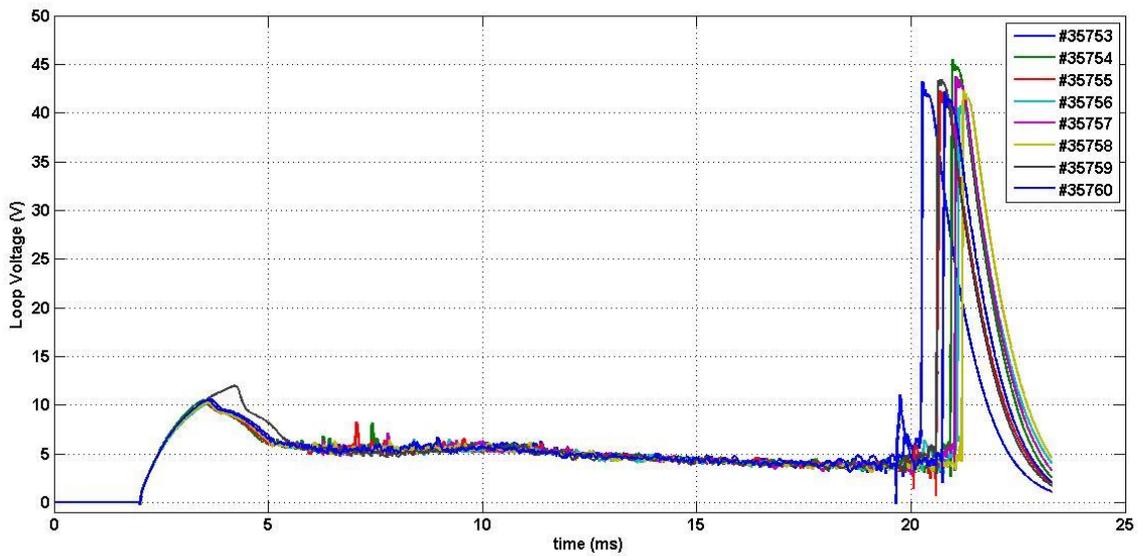

*Fig. 3* *Temporal evolution of loop voltage for 8 discharges under study*

During discharges under study plasma stabilization was not activated. Therefore, the plasma column moves inside discharge chamber (see the Fig.4).

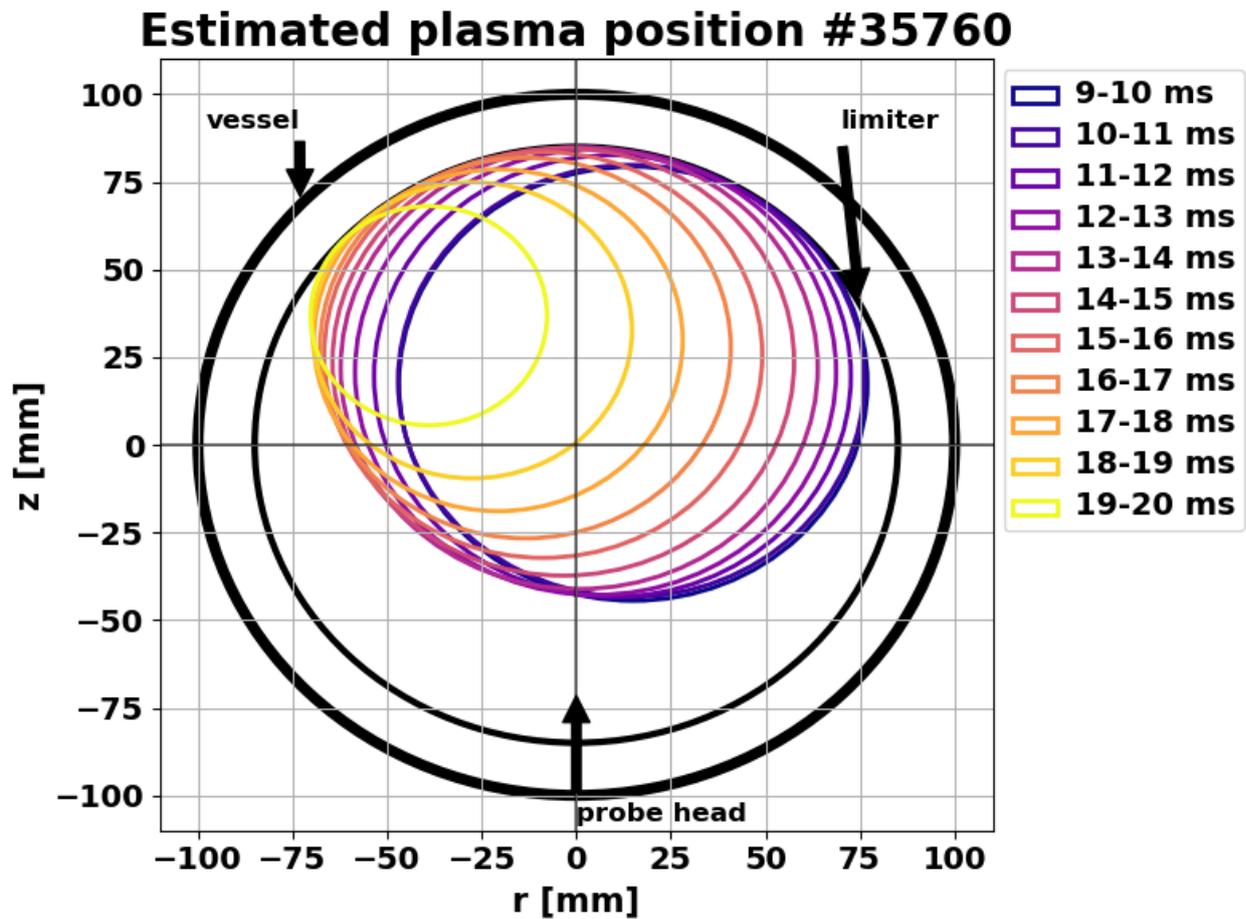

*Fig. 4 Plasma position in the discharge chamber of the GOLEM tokamak during a typical discharge*

During the 1 ms time window after 15 ms from the beginning of the discharge the plasma occupies only a small part of the discharge chamber and has contact with the poloidal limiter on its upper part at the high field side, as shown in Fig.5. The same figure demonstrates, once again, good reproducibility of the discharges in the GOLEM tokamak. It should be also mentioned that our measurements are generally performed in the SOL (scrape-off layer). Namely, the first pin of the left column on the rake probe (closest to the nose of the probe shaft), the results from which we present in this paper, is always outside the last closed flux surface (LCFS) and its radial position changes within the range r = 45÷80 mm.

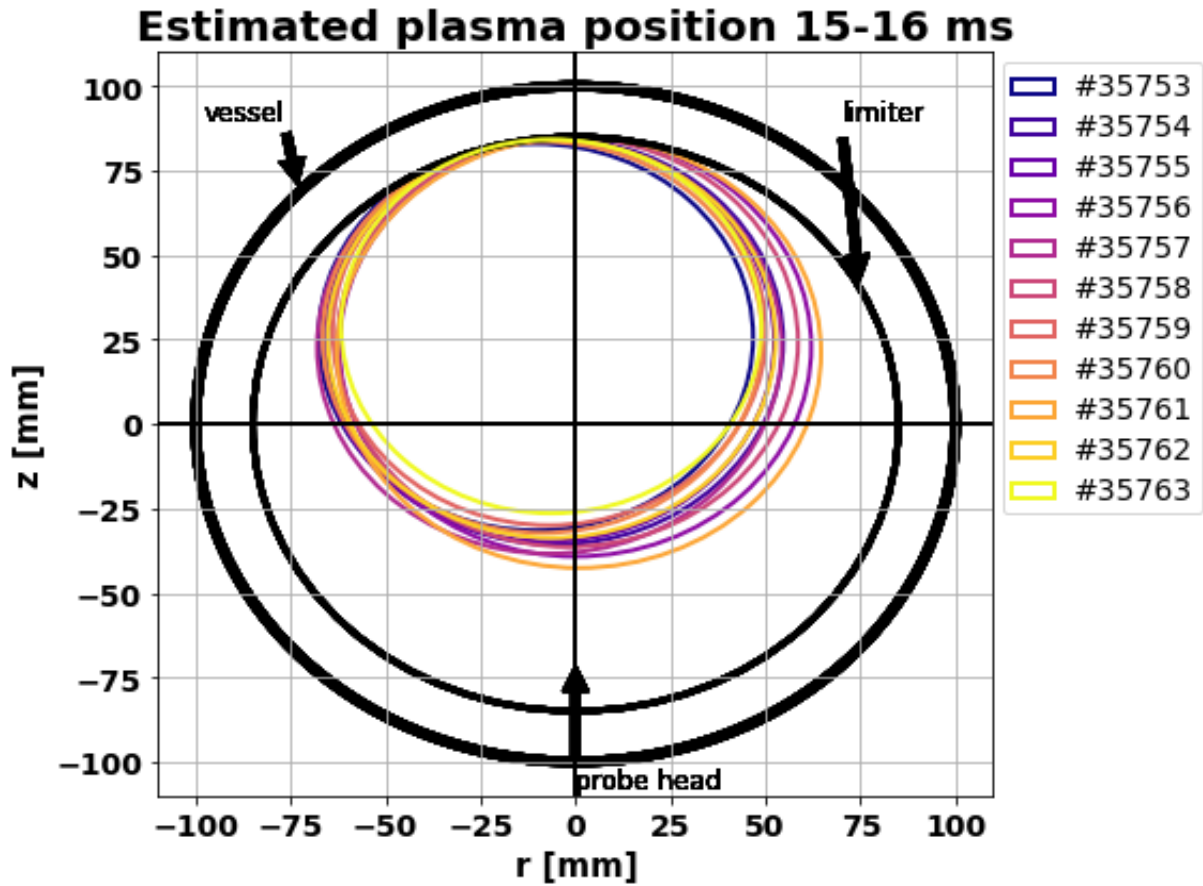

*Fig. 5 Plasma position in the discharge chamber of the GOLEM tokamak in 15-16 ms time window for the discharges under study*

## Experimental Results

Our study of the physical nature of plasma turbulent transport at the edge of the GOLEM tokamak involves statistical analysis as a first step. Radial dependence of $I_{sat}$ skewness is presented in Fig. 6, showing that $I_{sat}$ skewness is positive and generally increases radially. Positive skewness of plasma fluctuations and generally non-Gaussian behavior is quite common at the edge of tokamaks [2-6]. Positive skewness indicates the presence of coherent turbulent structures in tokamak edge plasma which propagate radially outwards. We have explained this by the physical picture described in [28-32]. Coherent structures are localized in a poloidal plane and are elongated along magnetic field line (mainly in toroidal direction). Plasma density and temperature inside them are higher than ambient plasma density and temperature. Coherent structures propagate radially outwards as a result of electric drift due to presence of poloidal electric and toroidal magnetic fields.

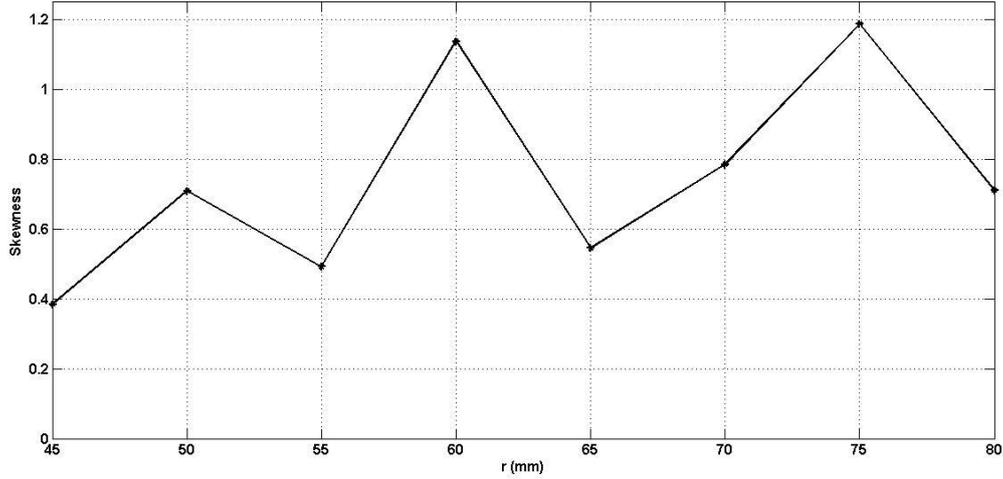

***Fig. 6*** *Radial profile of $I_{sat}$ skewness*

When coherent structures propagate radially outwards plasma density and temperature inside them decrease due to parallel losses (along the magnetic field line) and also deformation-spreading in the poloidal plane. At the same time plasma density and temperature inside coherent structures decrease more slowly than ambient plasma mean density and temperature (which also decrease radially outwards). Therefore, during the radial propagation plasma density and temperature inside coherent structures become increasingly higher than the same mean quantities for ambient plasma and when they pass the probe pin this gives rise to increasingly high amplitude positive $I_{sat}$ fluctuation (burst) and positive skewness. Thus, as the radial distance increases skewness also increases. This is what we observe in GOLEM – $I_{sat}$ skewness generally increases radially outwards (see the Fig. 6).

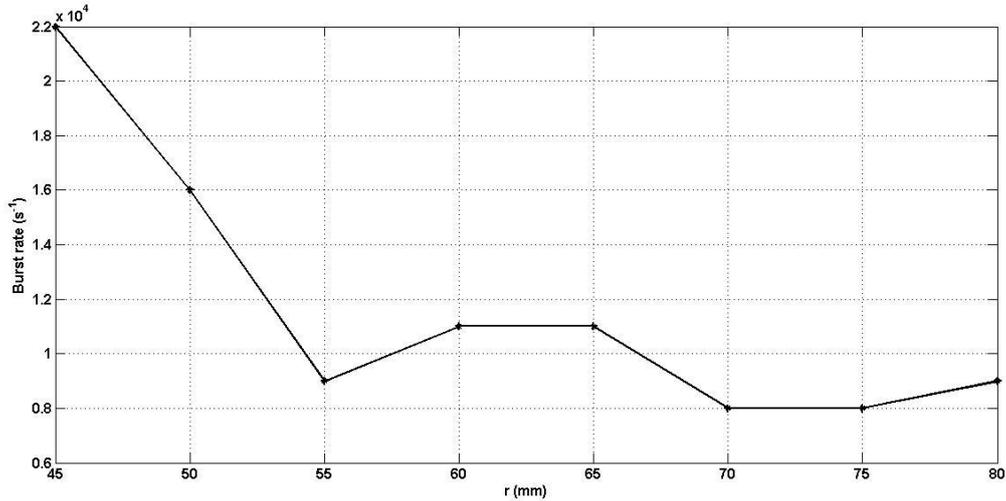

***Fig. 7*** *Radial profile of $I_{sat}$ average burst rate*

For a better understanding of plasma turbulent transport and coherent structure dynamics at the edge of tokamaks together with statistical characteristics we have successfully studied the radial dependence of $I_{sat}$ fluctuation temporal characteristics such as burst rate, inter-burst time and burst duration [28-32]. Bursts are selected using the threshold method. Here we use threshold of one standard deviation. Radial evolution of $I_{sat}$ average burst rate at the edge of the GOLEM tokamak is presented on Fig. 7.

We see that generally average burst rate decreases radially. The reason for this is that during radial propagation coherent structures decay as a result of parallel losses (along the magnetic field line) and also deformation-spreading in the poloidal plane. At the same time coherent structures have a distribution in size – the number of structures is decreasing as a function of their size. It is evident that larger structures propagate further and reach radially distant regions. Thus, as radial distance increases, a lower number of coherent structures reach this region and the average burst rate decreases [28-32].

The radial profile of the $I_{sat}$ average burst duration is presented on Fig. 8. The average burst duration increases radially. This happens due to deceleration of coherent structures during radial propagation and also their deformation-spreading in the poloidal plane – coherent structure becomes larger in the poloidal plane and moves more slowly. Consequently, as the radial distance increases, more time is necessary for coherent structures to cross the probe pin and the average burst duration increases [28-32].

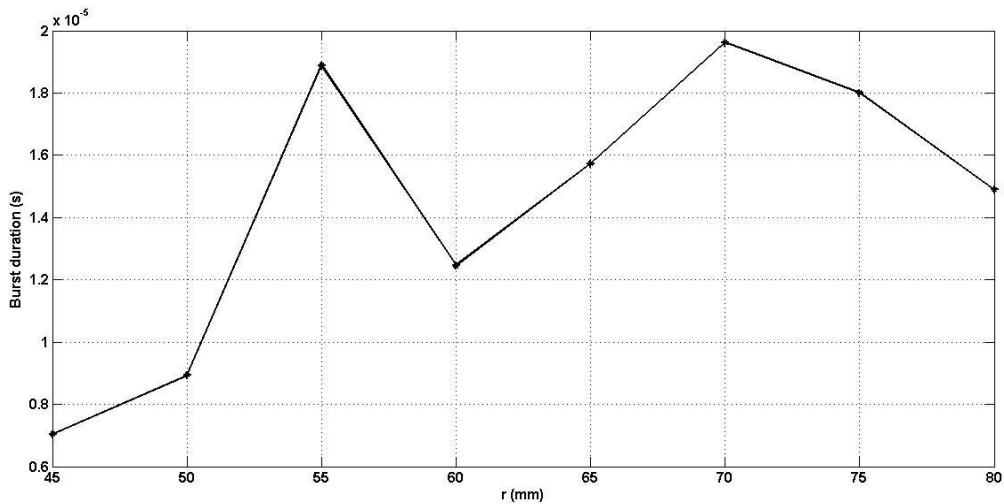

*Fig. 8* Radial profile of $I_{sat}$ average burst duration

## Discussion

The investigation of plasma turbulent transport at the edge of the CASTOR tokamak showed a similar radial dependence of statistical and temporal characteristics of $I_{sat}$ fluctuations [31]. Namely, skewness increased, average burst rate decreased and average burst duration increased in radial direction. But these radial profiles were much flatter than the same radial profiles from the GOLEM tokamak [31]. The reason is that radially elongated coherent structured – streamers governed plasma turbulent transport in the CASTOR tokamak [31, 33]. Formation of radially elongated structures was possible in the CASTOR tokamak due to high content of neutrals in edge plasma and high collisionality of plasma particles with neutrals [33]. Collisions damp zonal flows [34] and the formation of radially elongated structures - streamers becomes possible. In the paper [31] we performed a comparative analysis of fluctuation statistical and temporal characteristics of the CASTOR and Tore Supra tokamaks. In the case of Tore Supra tokamak these characteristics changed significantly in radial direction and profiles were not nearly-flat like in the CASTOR tokamak. In the framework of physical picture proposed by us it is evident, that when radial elongated streamer-like structures govern plasma radial transport, radial profiles of statistical and temporal characteristics will be nearly flat. But, if plasma transport is governed by blob-like structures with comparable poloidal and radial size, statistical and temporal characteristics of fluctuations will change significantly in radial direction.

We observe significant modification of statistical and temporal characteristics of plasma turbulent fluctuations at the edge of the GOLEM tokamak. Based on physical picture presented in the previous section it is evident that non-elongated – blob-like structures with comparable poloidal and radial size should be governing radial transport at the edge of the GOLEM tokamak.

It should be noted that a radial increase of $I_{sat}$ skewness was observed not only at the edge of the GOLEM tokamak, but also in Tore Supra [31], TCV [5] and TEXTOR tokamaks [6] as well as in the PISCES linear device [2]. In the JET tokamak a radial increase of $I_{sat}$ normalized fluctuation level was observed from 0.2 at the LCFS up to 0.6 in the SOL – 3 cm outside the LCFS [35]. All this demonstrates the universal nature of plasma turbulent transport.

The reason why blobs and not streamers are formed at the edge plasma of the GOLEM tokamak should be the following. Plasma temperature in the GOLEM tokamak is much lower than in the CASTOR tokamak. At the same time the cross-section of the plasma column is much smaller than the cross-section of the discharge chamber (in the CASTOR tokamak plasma column filled the entire poloidal cross-section) and the plasma has a small contact region with the poloidal limiter mostly at the upper high field side (see the Figures 4 and 5). For these reasons less neutrals are extracted from the limiter and their content in the plasma of the GOLEM tokamak should be much smaller than in the CASTOR tokamak. Therefore, collisionality in the GOLEM plasma should be low and thus zonal flows are not damped in the GOLEM tokamak and they do not allow the formation of radially elongated structures – streamers. Therefore, we have more blob-like structures at the edge of the GOLEM tokamak.

## Conclusion

Turbulent transport of non-stabilized plasma at the edge of the GOLEM tokamak was investigated. The study of the radial dependence of $I_{sat}$ fluctuation statistical and temporal characteristics together with a comparison with the results obtained during similar investigations performed on the CASTOR tokamak allowed to conclude that non-elongated – blob-like structures with comparable poloidal and radial size should govern plasma turbulent transport at the edge of the GOLEM tokamak.

## Acknowledgement

We dedicate this paper to the honorable memory of our colleague Dr. Jan Stöckel, who was the initiator of investigations in the field of plasma turbulent transport on the CASTOR and GOLEM tokamaks.